\documentclass[iop,apj]{emulateapj}
\usepackage{amsmath,amssymb,amstext}

\usepackage[breaklinks,colorlinks,citecolor=blue,linkcolor=magenta]{hyperref} 

\usepackage[all]{hypcap} 

\usepackage{amssymb}
\usepackage{amsmath}
\usepackage{graphicx}

\newcommand{\mbfu}{\mathbf{u}}

\newcommand{\bnabla}{{\mbox{\boldmath$\nabla$}}}
\newcommand{\bkappa}{{\mbox{\boldmath$\kappa$}}}

\usepackage{aas_macros}
\usepackage{natbib}
\shorttitle{Simulations of cosmic ray driven galactic winds}
\shortauthors{Ruszkowski, Yang, and Zweibel}

\begin{document}

\title{Global simulations of galactic winds including cosmic ray streaming}
\author{Mateusz Ruszkowski$^{1,2}$, H.-Y. Karen Yang$^{2,3}$, and Ellen Zweibel$^{4,5}$}
\affil{$^{1}$Department of Astronomy, University of Michigan, 1085 S University Ave, 311 West Hall, Ann Arbor, MI 48109}
\affil{$^{2}$Department of Astronomy, University of Maryland, College Park, MD 20742}
\affil{$^{4}$Department of Astronomy, University of Wisconsin-Madison, 475 N. Charter St, Madison, WI 53706}
\affil{$^{5}$ Department of Physics, University of Wisconsin-Madison, 1150 University Ave, Madison, WI 53706}
\altaffiltext{3}{{\it Einstein} Fellow}
\email{mateuszr@umich.edu (MR), hsyang@astro.umd.edu (KY), zweibel@astro.wisc.edu (EZ)}

\begin{abstract}
Galactic outflows play an important role in galactic evolution. Despite their importance, a detailed understanding of the physical mechanisms responsible for the driving of these winds is lacking. In an effort to gain more insight into the nature of these flows, we perform global three-dimensional magneto-hydrodynamical simulations of an isolated Milky Way-size starburst galaxy. We focus on the dynamical role of cosmic rays injected by supernovae, and specifically on the impact of the streaming and anisotropic diffusion of cosmic rays along the magnetic fields. We find that these microphysical effects can have a significant effect on the wind launching and mass loading factors depending on the details of the plasma physics. Due to the cosmic ray streaming instability, cosmic rays propagating in the interstellar medium scatter on self-excited Alfv{\'e}n waves and couple to the gas. When the wave growth due to the streaming instability is inhibited by some damping process, such as the turbulent damping, the cosmic ray coupling to the gas is weaker and their effective propagation speed faster than the Alfv{\'e}n speed. Alternatively, cosmic rays could scatter from ``extrinsic turbulence'' that is  driven by another mechanism. We demonstrate that the presence of moderately super-Alfv{\'e}nic cosmic ray streaming enhances the efficiency of galactic wind driving. Cosmic rays stream away from denser regions near the galactic disk along partially ordered magnetic fields and, in the process, accelerate more tenuous gas away from the galaxy. For cosmic ray acceleration efficiencies broadly consistent with the observational constraints, cosmic rays reduce the galactic star formation rates and significantly aid in launching galactic winds.
\end{abstract}
 
\maketitle

\section{Introduction}
Most galaxies contain only a small fraction of baryons compared to the cosmological average (e.g., \citet{bell2003}). Models that rely on matching observed galaxy luminosity functions to simulated halo mass functions (e.g., \citet{conroy2006, guo2010}) reveal that about 20\% of baryons are accounted for in $L_{*}$ galaxies and this fraction declines for both more and less massive galaxies. These missing baryons either did not fall into the potential wells of forming protogalaxies \citep{anderson2010} or were expelled as a result of feedback processes operating during galaxy formation. In the case of galaxies more massive than $L_{*}$ this feedback is likely dominated by AGN activity (e.g., \citet{birzan2004, croton2006}), while in less massive objects the outflows could be a result of starburst activity (e.g., \citet{larson1974, white1978, dubois2010}). Galactic winds are almost universally observed in galaxies that have recently experienced intense star formation episodes \citep{veilleux2005}. These outflows drive the gas out of galaxies at rates ranging from 1\% of the star formation rate (SFR) to ten times higher than the observed star formation rates \citep{bland-hawthorn2007}. Thus, galactic winds play an essential role in galactic evolution. Galactic models that do not include feedback processes suffer from overpredicting the amount of baryons and star formation rates \citep{crain2007, stinson2013}. However, despite their importance, detailed understanding of the physical mechanisms responsible for the driving of these winds is lacking. Some of the feedback mechanisms invoked to drive the winds include thermal feedback from supernovae \citep{joung2009} and radiation pressure \citep{murray2005, murray2011, hopkins2012}. \\
\indent
Recently, the role of cosmic rays (CR) in galactic feedback began to receive significant attention (e.g., \citet{breitschwerdt1991, everett2008, socrates2008}). VERITAS gamma-ray observations of a starburst galaxy M82 \citep{veritas2009} suggest that cosmic ray energy density exceeds that seen in the Milky Way by orders of magnitude. {\it Fermi} observations of M82 and NGC 253 also imply high CR energy densities \citep{paglione2012}. Significant outflows have been inferred in these objects (e.g., \citet{ackermann2012, yoasthull2013, bolatto2013}). 
In the Milky Way, the CR energy density is comparable to that in turbulence and magnetic fields. These CR were likely generated via shock acceleration in supernova remnants (e.g., \citet{Blandford1987}; see \citet{caprioli2015} for a recent review) and winds from massive young stars \citep{bykov2014}.\\
\indent
The impact of CR on wind launching has been studied using a variety of approaches. \citet{breitschwerdt1991} considered one-dimensional steady wind models that included cosmic ray streaming along large scale coherent magnetic fields and pressure forces due to the scattering of CR on self-excited Alfv{\'e}n waves. They demonstrated that this mechanism can efficiently accelerate the gas from the disk. This work was extended by \citet{everett2008}, who studied the impact of combined CR and thermal pressure in the Milky Way and concluded that CR pressure is essential for wind driving. These papers presented proof-of-concept results that highlighted the importance of CR in wind launching. \\
\indent
Three-dimensional magnetohydrodynamical (MHD) simulations of zoom-in regions of galactic disks that included anisotropic CR diffusion and a sophisticated treatment of chemical network to compute gas cooling were recently presented by \citet{girichidis2016}, who demonstrated that CR can drive a smooth cold wind and thicken galactic disks provided that the CR are coupled to the gas. In a closely related extension of this work, \citet{peters2015} presented a study of the X-ray emission from such flows.\\
\indent
The dynamical role of CR in launching the winds was also studied via purely hydrodynamical global simulations by \citet{uhlig2012}, who considered CR streaming; \citet{booth2013}, who included CR energy losses and isotropic CR diffusion; and \citet{salem2014} who performed simulations for a range of isotropic CR diffusivities. Global MHD simulations were presented by \citet{hanasz2013}, who considered a non-radiative model for a single value of magnetic-field-aligned CR diffusion. These simulations with their progressive level of sophistication, demonstrated that CR play an important role in wind launching. \\
\indent
In an effort to gain more insight into the nature of CR driven winds, we perform global three-dimensional MHD simulations of an isolated starbursting galaxy. Our simulations include magnetic fields, radiative cooling, self-gravity, star formation, dynamical role of CR injected by supernovae, and CR transport processes -- anisotropic diffusion and cosmic ray streaming along the magnetic fields.  We thus extend some of the physics treatments considered in earlier global simulations of CR-driven winds. Our treatment is unique in part because of the modeling of transport processes that, in addition to anisotropic CR diffusion, accounts for CR streaming mediated by the streaming instability. We demonstrate that CR provide efficient feedback that reduces star formation rates and significantly aid in launching galactic winds, and that the above plasma processes are essential for the efficient CR wind driving. 
The other inputs to our models - star formation, stellar energy input, gravitational potential, etc. follow commonly used prescriptions, a deliberate choice to facilitate comparison with other models and isolate the role of CR. \\
\indent
The outline of the paper is as follows. In Section 2 we discuss the physical process included in the simulations, describe our model and parameter choices in the performed simulations. Results are presented in Section 3 and conclusions in Section 4.
\section{Methods}
We solve the MHD equations including CR advection, dynamical coupling between CR and the thermal gas, CR streaming along the magnetic field lines and the associated heating of the gas by CR, anisotropic CR diffusion, self-gravity of the gas, and radiative cooling
\begin{align}
\frac{\partial \rho}{\partial t} + \bnabla \cdot (\rho {\bf u}_{g})  &=  0,\\
\frac{\partial \rho {\bf u}_{g}}{\partial t} + \nabla \cdot \left( \rho {\bf u}_{g}{\bf u}_{g}- \frac{{\bf B}{\bf B}}{4\pi} \right) + \bnabla p_{\rm tot} &=  \rho {\bf g} + {\dot p}_{\rm SN},\\
\frac{\partial {\bf B}}{\partial t} - \bnabla \times ({\bf u}_{g}\times {\bf B})  & =  0, \label{eq:ind}\\
\frac{\partial e}{\partial t} + \bnabla \cdot \left[ (e+p_{\rm tot}){\bf u}_{g} - \frac{{\bf B}({\bf B} \cdot {\bf u}_{g})}{4\pi} \right]  &=  \rho {\bf u}_{g}\cdot {\bf g}\nonumber \\
-\nabla \cdot {\bf F}_{\rm c} + \bnabla\cdot ({\bkappa}\cdot\bnabla e_{c}) - {\cal C} & +{\cal H}_{\rm c}+{\cal H}_{\rm SN},   \label{eq:4} \\
\frac{\partial e_{\rm c}}{\partial t} + \bnabla \cdot (e_{\rm c} {\bf u}_{g}) = -p_{\rm c}\bnabla\cdot {\bf u}_{g} & - {\cal H}_{\rm c}+{\cal H}_{\rm SN}\nonumber  \\
  -\nabla \cdot {\bf F}_{\rm c}  & +\bnabla\cdot ({\bkappa}\cdot\bnabla e_{c}),    \label{eq:5} \\
\Delta\phi =  4\pi G\rho_{b} &
\end{align}
where $\rho$ is the gas density, $\rho_{b}$ is the total baryon density, ${\bf u}_{g}$ is the velocity, ${\bf B}$ is the magnetic field, ${\bf g}=-\nabla\phi + {\bf g}_{\rm gal}$ is the gravitational field (including the contribution from self-gravity of the gas $-\nabla\phi$, stars, and other components ${\bf g}_{\rm gal}$ described below (see Section 2.4)), ${\dot p}_{\rm SN}$ is the rate of momentum injection associated with stellar mass loss due to stellar winds and SN, $e_{\rm c}$ is the specific CR energy density, and $e=0.5\rho {\bf u}_{g}^2 + e_{\rm g} + e_{\rm c} + B^2/8\pi$ is the total energy density, ${\cal C}$ is the radiative cooling energy loss rate per unit volume (see Section 2.1), ${\bf F}_{\rm c}$ is the CR flux due to streaming relative to the gas, ${\cal H}_{\rm c}$ is the CR heating due to the streaming instability (see Section 2.2 for the discussion of CR streaming), and ${\cal H}_{\rm SN}$ represents heating due to supernovae (see section 2.3). The total pressure is $p_{\rm tot} = (\gamma_{g} -1)e_{\rm g} + (\gamma_{c} -1) e_{\rm c} + B^2/8\pi$, where $e_{\rm g}$ and $e_{\rm c}$ are the specific thermal energy density of the gas, $\gamma_{g}=5/3$ is the adiabatic index for ideal gas, and $\gamma_{c}=4/3$ is the effective adiabatic index of CR fluid. \\
\indent 
We solve the above equations using the adaptive mesh refinement (AMR) MHD code FLASH4.2. We employ the directionally unsplit staggered mesh solver \citep{Lee09, Lee13}. This solver is based on a finite-volume, high-order Godunov scheme and utilizes a constrained transport method to enforce divergence-free magnetic fields.

\subsection{Radiative cooling}
We include radiative cooling losses using the Sutherland \& Dopita cooling function \citep{SutherlandDopita93} extended down to 300 K following the cooling function of \citet{DalgarnoMcCray72}. We also include mean molecular weight variations caused by the changes in the gas ionization state. This is accomplished by interpolating the \citet{SutherlandDopita93} tables for temperatures $T\ge10^{4}$K (for simplicity, for $T<10^{4}$K we take the largest mean molecular weight from their table). Whenever the cooling time falls below the hydrodynamical Courant timestep, we employ the subcycling method \citep{anninos97, proga03} in order to accelerate the computations.

\subsection{Cosmic ray streaming}
Implementation of streaming in our wind models, one of the major accomplishments of this paper, has given us the freedom to explore
several different treatments of cosmic ray transport, which we classify as ``self confinement" and ``extrinsic turbulence" \citep{zweibel2013}. 

According to the self confinement picture, CR gyrate around magnetic fields with frequency $\Omega_{o}/\gamma$ and gyroradius $r_{g}=\gamma c/\Omega_{o}$, where $\Omega_{o}$ is the usual non-relativistic cyclotron frequency.  An Alfv{\'e}n wave can strongly interact with a CR when the $k_{||}=1/(\mu r_{g})$ resonance condition is met, where $\mu$ is the cosine of the CR pitch angle, and $k_{||}$ is the wave vector of the CR propagating parallel to the local magnetic field. This condition arises because, for much larger wave vectors,   small scale fluctuating Lorentz forces acting on CR essentially cancel out, while for much smaller wave vectors CR experience only large scale magnetic field and slide along the fields without getting scattered. Thus, CR interact particularly strongly with waves which meet the gyroresonance condition.

It was shown by  \citep{kulsrud1969, wentzel1974} that Alfv{\'e}n waves are amplified by gyroresonant scattering if the CR have a positive streaming anisotropy in the wave frame. That is, super-Alfv{\'e}nic streaming amplifies Alfv{\'e}n waves traveling in the same direction as the streaming, and damps waves traveling in the opposite
direction. Scattering by these amplified waves drives the CR toward a state of marginally stable anisotropy. If there were no
damping mechanism to counteract excitation by streaming CR, CR would stream at the Alfv{\'e}n speed. However, because
some form of damping is generally present, streaming is always at least slightly super-Alfv{\'e}nic, becoming more so as the strength of the damping increases. 

The waves which scatter the CR cause spatial diffusion along the magnetic fieldlines. \citet{skilling1971} showed that when the waves are marginally stable, the diffusion coefficient $\kappa$ cannot be chosen freely, but must be adjusted  such that  the cosmic ray anisotropy is consistent with marginal stability. This extra constraint reduces the order of eqns. (\ref{eq:4}) and (\ref{eq:5}) below what the diffusion term formally implies. 

In the extrinsic turbulence picture, the waves which scatter the CR are not driven by the CR themselves, but are part of
a turbulent cascade (we will continue to assume that the waves are Alfv{\'e}n waves). In this case, the waves drive the CR toward
isotropy in a frame traveling with the weighted mean velocity \citep{skilling1975}
\begin{equation}\label{uf}
\mbfu_f\equiv\mbfu_A\frac{\nu_{+}-\nu_{-}}{\nu_{+}+\nu_{-}},
\end{equation}
where $\nu_{\pm}$ denote the scattering frequencies by waves in the $\pm$ directions. If the $\pm$ waves have the same spectra (``balanced
turbulence"),  the CR are just advected with the thermal fluid.  The diffusion coefficient $\kappa$ can represent not only gyroresonant scattering,
but also large scale but unresolved fluctuations in the background magnetic field $B$. Since our global simulations are far from resolving 
known turbulent scales in the interstellar medium, it makes sense to parameterize  cosmic ray transport along randomly wandering fieldlines as
spatial diffusion.

We can model transport in both the self confinement and extrinsic turbulence pictures by introducing a parameter $f$ such that the advection
speed velocity $\mbfu_c$ in eqns. (\ref{eq:4}) and (\ref{eq:5}) is
\begin{equation}\label{fdef}
\mbfu_c= \mbfu_g +f\mbfu_A.
\end{equation}
For the self confinement picture, we follow \citet{kulsrud1971} and \citet{wiener2013} and represent the effect of wave damping by
choosing $f > 1$ and dropping the spatial transport term $\kappa$ unless we wish to model spatial transport by fieldline wandering.
A number of damping mechanisms may operate, e.g., ion-neutral damping, non-linear Landau damping, or turbulent damping by background MHD turbulence. Let us consider turbulent damping as an example. In this case, the Alfv{\'e}n waves excited by CR collide with turbulence generated wave packets propagating in the opposite direction and dissipate \citep{farmer2004}. The balance of wave growth and decay rates leads to the following value of $f$ \citep{wiener2013}
\begin{equation}
f=\left(1 + 8.0\frac{B_{\mu G}^{1/2}n_{i,-2}^{1/2}}{L_{\rm mhd, 10}^{1/2}   n_{c,-10}}   \gamma_{3}^{n-3.5}10^{2(n-4.6)}\right),
\end{equation}
where $n_{i,-2}=n_{i}/(10^{-2}{\rm cm}^{-3})$ is the ion number density,  $n_{c,-10}=n_{c}/(10^{-10}{\rm cm}^{-3})$ is the CR number density, $L_{\rm mhd, 10}=L_{\rm mhd}/(10 {\rm pc})$ is the lengthscale at which turbulence is driven with velocity comparable to Alfv{\'e}n speed $u_{A}$, $\gamma_{3}=\gamma/3$ is the mean CR Lorentz factor, and $n$ is the slope of the CR distribution function in momentum ($n=4.6$ in the Milky Way). The ratio $(B_{\mu G}/L_{\rm mhd, 10})^{1/2}$ is essentially a free parameter in our model as there is considerable uncertainty in the determination of the initial strength of the magnetic field and turbulence injection scale.  As an example, for plausible values near or above the disk $L_{\rm mhd, 10} = 1$ (\citet{iacobelli2013} and references therein), $B_{\mu G} = 3$ \citep{beck2009}, $n_{i,-2}= 0.5$ and $n_{c,-10} = 3$, we get 
$f = 4.3$, i.e., a moderate boost in the effective streaming speed. If we include diffusion, we interpret it as due to fieldline random walk.
Note that this estimate assumed  constant spectral shape (constant $n$). A slight steepening of the spectrum due to the energy dependent streaming would somewhat increase the effective $f$. Moreover, other damping mechanisms, such as ion-neutral damping, may further reduce CR coupling to the gas and help to boost the streaming speed. In general, in cool dense clumps the coupling of CR to the gas will be weak and the transport will proceed locally at larger speeds. Such processes are best studied using small-scale simulations that include detailed decoupling physics. We will present results from such studies in a forthcoming publication. In the current work, given the approximate nature of these considerations, we simply set $f$ to a constant in this preliminary study. This allows us to build some intuition and assess the impact of the effective streaming speed on the results by varying $f$ between simulations. As we show below, a moderate boost in the streaming speed helps to launch the wind.

For the extrinsic turbulence picture, we simply take $f < 1$, and let $\kappa$ be a free parameter.

As CR stream down the pressure gradient and scatter on the the MHD waves they experience an effective drag force that heats the gas. In the self confinement picture this heating term is  $-{\bf u}_{A}\cdot\nabla p_{c}$,  not $-{\bf u}_{c}\cdot\nabla p_{c}$. This is because the frictional force results from the coupling of CR to the gas, and the super-Alfv{\'e}nic streaming is due to the reduction of this coupling. The transfer of energy from CR to the gas is thus facilitated only by that portion of the streaming that is due to the MHD waves, i.e., by the Alfv{\'e}n speed. This is shown explicitly in the Appendix.  In the extrinsic turbulence picture, i.e., $f\le 1$, the heating rate is  $- f{\bf u}_{A}\cdot\nabla p_{c}$ \\
\indent
Cosmic ray advection and coupling to the thermal gas is included in the simulations using the modules described in our earlier work \citep{y12,y13}. We now extend this treatment to include cosmic ray streaming following the method of \citet{sharma09}. The CR streaming flux is given by ${\bf F}_{\rm c} = (e_{\rm c}+p_{\rm c}){\bf u}_{s}$, where
${\bf u}_{s}=-{\rm sgn}(\hat{{\bf b}}\cdot\nabla e_{\rm c}){\bf u}_{A}$ is the streaming velocity, where ${\bf u}_{A}$ is the Alfv{\'e}n velocity and $\hat{{\bf b}}$ is the magnetic direction vector. In the presence of turbulent dumping, the last two terms in Eq. (5), i.e., the streaming and diffusion terms, can be approximated by combining them into a single term that mathematically behaves just as the regular streaming term in the absence of significant damping but with the Alfv{\'e}n speed replaced by ${\bf u}_{D}$.\\
\indent
The $-\bnabla\cdot {\bf F}_{c}$ term on the right hand side of Eqs. (4) and (5) describes the streaming of CR and is very similar in appearance to the usual advection term. However, unlike the regular advection term, the streaming term also depends on the gradient in CR energy density. Near the local extrema in the CR energy density distribution, the streaming flux is discontinuous and this term varies infinitely fast, which leads to a prohibitively small code timestep. In order to address this problem, we regularize the streaming flux by 
${\bf F}_{\rm c} = -(e_{\rm c}+p_{\rm c}){\bf u}_{A}{\rm tanh}(h_{\rm c}\nabla e_{\rm c}/e_{\rm c})$, where $h_{c}$ is a free (regularization) parameter. With this regularization, the streaming term becomes diffusive in nature near the extrema in CR energy density and the singularity is removed. In order to further accelerate computations we subcycle over the CR  streaming term. Since each AMR block contains four guard cells, we can only subcycle up to four times. The new timestep constraint becomes 
\begin{eqnarray}
dt = 1.8\times 10^{11}\left(\frac{f_{\rm cfl}}{0.8}\right)\left(\frac{N}{4}\right)\left(\frac{dx}{100\:{\rm pc}}\right)^{2}  \nonumber \\
\left(\frac{u_{D}}{200\:{\rm km/s}}\right)^{-1}\left(\frac{h_{\rm c}}{10\:{\rm kpc}}\right)^{-1}{\rm s}, 
\end{eqnarray}
\noindent
where $f_{\rm cfl}$ is the diffusive Courant number and $N$ is the number of subcycles over the streaming term. We present tests of the streaming module in the Appendix. In the galaxy simulations including streaming we use $h_{\rm c}=10$ kpc. We verified that the results do not depend on the choice of $h_{\rm c}$ by performing more CPU-intensive test runs for two times higher values of $h_{\rm c}$. We impose a limit on $u_{D}$ both in the flux terms and the CR heating terms. Our production runs use the ceiling of 200 km/s. However, we also performed more expensive simulations for two times higher ceiling values and obtained essentially identical results.

\subsection{Star formation and supernova feedback}
Star formation is based on the approach of \citet{CenOstriker92} (see also \citet{TaskerBryan06, salem2014, Li15, bryan2014}). Star formation is triggered when the following conditions are simultaneously met: (i) gas density exceeds a threshold value of $1.67\times 10^{-23}$g cm$^{-3}$ \citep{gnedin2011, agertz2013}, (ii) cell gas mass exceeds local Jeans mass, (iii) $\nabla\cdot{\bf v} < 0$, and (iv) gas temperature reaches the floor of the cooling function or the cooling time becomes shorter than the dynamical time $t_{\rm dyn}=(3\pi/(32G\rho_{b}))^{1/2}$, where $\rho_{b}$ is the total density in the baryonic component. Once these conditions are met, a star particle is formed and its mass is increased over time at the rate of $\dot m = m_{*}(\Delta t/\tau^{2})\exp(-\Delta t/\tau)$, where $\Delta t$ is the time since the formation of the star particle, $\tau = \max (t_{\rm dyn}, 10{\rm Myr})$, and $m_{*}$ is the final mass of the star particle. This mass is computed according to $m_{*}=\epsilon_{\rm SF} (dt/t_{\rm dyn})\rho dx^{3}$, where $dt$ is timestep, $dx$ is the size of the cell containing the particle, and $\epsilon_{\rm SF}=0.05$ is the efficiency of star formation. We set a minimum $m_{*, {\rm min}}=10^{5} M_{\odot}$ in order to prevent the code from slowing down due to excessively large number of star particles. Whenever $m_{*}<m_{*, {\rm min}}$ new stars characterized by the final mass of $0.8\rho dx^{3}$ are allowed to form with a probability of $m_{*}/m_{*, {\rm min}}$.\\
\indent
We introduce stellar feedback by adding gas mass at the rate $f_{*}{\dot m}$, thermal energy $(1-f_{\rm cr})\epsilon_{\rm SN}{\dot m}c^{2}$, and cosmic ray energy $f_{\rm cr}\epsilon_{\rm SN}{\dot m}c^{2}$  to the gas in the cell containing the star particle. During this time-dependent feedback we ensure that the mass added to the gas is subtracted from the stellar particle mass in order to ensure conservation of total baryon mass. The mass injection is associated with stellar mass loss due to SN and stellar winds, and the added mass is assigned the velocity of the parent star. We assume fraction of returned mass  $f_{*}=0.25$ and $\epsilon_{\rm SN}=10^{51}{\rm erg}/(M_{\rm sf}c^{2})$, where $\epsilon_{\rm SN}$ specifies the amount of energy injected by one supernova per $M_{\rm sf}$ of mass in newly formed stars. The considered values of $f_{\rm cr}$ and $M_{\rm sf}$ are listed in Table 2. In our fiducial run we use $M_{\rm sf}=100M_{\odot}$ for the SN efficiency \citep{guedes2011,hanasz2013}. \\
\indent
The acceleration efficiency $f_{\rm cr}$ is poorly constrained. A number of studies reported values ranging from $\sim 0.1$ to $\sim 0.4$ \citep{helder2013,morlino2015,hess2016}. However, both smaller (Hewitt, priv. comm.) and larger \citep{kang2006,ellison2010} values have been inferred. Detailed constraints on the acceleration efficiency can be obtained in some specific cases, notably the well-studied Tycho SN remnant. Sophisticated theoretical models that utilize morphological and spectral data have been applied to this case \citep{morlino2012,berezhko2013,slane2014,caprioli2015}. Results based on these studies concluded that the CR efficiency is about 0.15 in this case. Global energetics arguments suggest that SN acceleration efficiencies between $\sim$0.05 and $\sim$0.15 are sufficient to explain current CR luminosity of the Milky Way \citep{hillas2005}. In addition to SN, fast winds from massive young stars can be an important source of CR \citep{bykov2014} thus increasing the net efficiency of CR acceleration. For the CR acceleration efficiency we use $f_{\rm cr}=0.15$ in our fiducial model, but we also include cases for $f_{\rm cr}=0.1$ and $f_{\rm cr}=0.3$.\\
\indent
We note that these are by no means unique choices of parameters or methods. We have experimented with different parameter choices and star formation and feedback prescriptions (e.g., different efficiencies, critical density thresholds, feedback delay times, spatially-distributed feedback from SNe) and we defer the study of the impact of these factors to a future publication. We note that in the current work we specialize to one type of feedback method and focus on the relative comparison between the models with and without CR physics in order to isolate the impact of CR streaming on the wind driving.
\begin{table}
  \caption{Parameters of the galaxy model}
  \label{tab:table1}
  \begin{center}
    \leavevmode
    \begin{tabular}{ll} \hline \hline              
  Halo                  \\ \hline 
   $M_{200}$ &   $10^{12}$M$_{\odot}$      \\
   $c$ &  12                   \\ \hline\hline
   Disk                 \\ \hline
   $r_{o}$  & 3.5 kpc \\
   $z_{o}$ & 0.325 kpc \\
   $\rho_{o}$ & $4.06\times 10^{-23}$g cm$^{-3}$\\
   $T$ & $10^{4}$K\\ \hline
  \multicolumn{2}{l}{}                            
   \end{tabular}
  \end{center}
\end{table}
\subsection{Galaxy model and initial conditions}
In order to set up the gas density in the galactic disk in the vertical direction, we follow the equilibrium solution for a stratified isothermal and self-gravitating system \citep{spitzer42, camm50,salem2014}. The gas density is given by 
\begin{equation}
\rho (r,z) = \rho_{o}e^{-\frac{r}{r_{o}}}{\rm sech}^{2}\left(\frac{z}{2z_{o}}\right).
\end{equation}
within $|z|<6z_{o}$ or $r<6r_{o}$, where $r$ is the distance from the rotation axis of the galaxy and $z$ is the distance along the vertical axis. The magnetic field in this region is toroidal and set up using the vector potential approach to ensure vanishing divergence of the magnetic field.  A constant magnetic field strength $B_{o}$ is assumed. Thus, in the initial state, the effective CR transport in the vertical direction vanishes. The magnetic pressure is in general not negligible compared to gas pressure in the disk in the initial state, and becomes more important as the gas cools down. For $|z|\ge6z_{o}$ or $r\ge 6r_{o}$, we set $\rho = 10^{-31}$g cm$^{-3}$ and $B=0$. The initial temperature is set to $10^{4}$K throughout the computational volume.
Total gas mass in the disk is approximately given by
\begin{equation}
M_{\rm gas}(<r) = 8.0\pi\rho_{o} z_{o} r_{o}^{2}e^{-\frac{r}{r_{o}}} \left( e^{\frac{r}{r_{o}}} -\frac{r}{r_{o}}-1.0 \right), 
\end{equation}
\noindent
which for $r=6r_{o}$ corresponds to $M_{\rm gas} = 6\times 10^{10}$M$_{\odot}$.\\
\indent
The static dark matter mass distribution is described using the NFW model \citep{navarro97}
\begin{equation}
M(<r)=M_{200}\frac{\ln(1+x)-x/(1+x)}{ \ln(1+c)-c/(1+c)},
\end{equation}
\noindent
where $x=r\:c/r_{200}$, $r_{200}=(M_{200}/[(200\rho_{\rm crit})4\pi/3)]^{1/3}$, and $\rho_{\rm crit}$ is the critical density of the Universe. Key model parameters are summarized in Table 1. We assume that the gas posses only a circular component given by $v_{\rm orb}(r)=(GM_{\rm tot}(<r)/r)^{1/2}$, where $M_{\rm tot}$ is the sum of the dark matter  and gas mass enclosed within a sphere of radius $r$.\\
\indent
We applied outflow boundary conditions to all sides of the domain. We used static non-uniform mesh refinement with 8 levels of refinement corresponding to the effective resolution of $1024^{3}$ elements and the linear resolution of 195 pc. The central portion of the domain containing the disk was fully refined. This resolution is comparable to that adopted in (\citet{dubois2010}; 150 pc), essentially the same as that in \citet{hanasz2013} who consider $512^{3}$ resolution in a box half the linear size, but worse than that in pure hudrodynamical simulations of \citet{salem2014}. Due to the stringent numerical demands imposed by the streaming physics and large Alfv{\'e}n speeds, our effective resolution does not exceed the above level in long runs. However, we also made an effort to perform a limited number of shorter resolution tests for the resolution of 98 pc in the streaming case. We observed that, while this resulted in the increase of the instantaneous SFR at early times and a slightly lower rates at later times, this improved resolution had a lesser effect on the integrated star formation and had practically no impact on the mass loading factor.  
\begin{table}
  \caption{List of simulations}
  \label{tab:table2}
  \begin{center}
    \leavevmode
    \begin{tabular}{ccccc} \hline \hline              
  $f_{\rm cr}$ & $M_{\rm sf}$ [M$_{\odot}$/SN] & B$_{o}$ [$\mu$G] & $f$ & $\kappa_{||}$ [cm$^{2}$s$^{-1}$]                 \\ \hline 
  0.10  & 100 & 1 & 4 & 0  \\
  0.15  & 100 & 1 & 4 & 0  \\
  0.30  & 100 & 1 & 4 & 0  \\
  0.15  & 100 & 3 & 4 & 0  \\
  
  0.10  & 100 & 1 & 0 & 0  \\
  0.10  & 100 & 1 & 1 & 0  \\
  0.10  & 100 & 1 & 8 & 0  \\

  0.10  & 185 & 1 & 0 & 0\\
  0.10  & 185 & 1 & 1 & 0\\  
  0.10  & 185 & 1 & 3 & 0\\
  0.10  & 185 & 1 & 0 & 3$\times 10^{27}$\\
  0.10  & 185 & 1 & 0 & 10$^{28}$\\ \hline
  \multicolumn{5}{l}{}                           
   \end{tabular}\par
   \end{center}
From left to right the columns list the values of the cosmic ray acceleration efficiency $f_{\rm cr}$, supernova feedback efficiency $M_{\rm sf}$ (in units of M$_{\odot}$ in newly formed stars per supernova, initial magnetic field B$_{o}$, effective streaming speed $f$ in units of the Alfv{\'e}n speed, and parallel cosmic ray diffusivity $\kappa_{||}$. 
\end{table}

\section{Results}
The list of runs corresponding to different parameter sets is shown in Table 2. All simulations include radiative cooling, self-gravity, magnetic fields, CR, and star formation. We begin the discussion of our results by considering two reference cases shown in Figure 1. From left to right the panels show cross-sections through the distribution of gas density, specific energy density in CR, and the vertical component of the magnetic field. The cross-sectional plane is perpendicular to the $x$ axis and passes through the center of the computational domain. In all panels the galaxy is seen edge-on.  All snapshots were taken at 400 Myr since the beginning of the simulation. Top row corresponds to $f=0$ (i.e., no CR streaming) and bottom to $f=4$. All panels correspond to the SN feedback efficiency of 100$M_{\odot}$/SN and CR acceleration efficiency $f_{\rm cr}=0.1$. \\
\indent
Figure 1 demonstrates that cosmic ray streaming has a dramatic effect on the wind launching from the galaxy. This is evidenced by the extent of the gas below and above galactic mid plane shown in the left column. In the case without streaming the characteristic vertical scaleheight of the gas density distribution is much smaller compared to the very extended gas distribution in the case that includes cosmic ray streaming. We note in passing that the mere presence of CR does have a moderate impact on the disk appearance even in the absence of streaming. Compared to pure MHD run (not shown), such a case results in a slightly puffed up disk. The differences in the gas density distributions are mirrored by those in the distribution of the specific energy density of CR (middle column). While the maximum height reached by CR above the disk midplane is comparable between these two cases, the lateral extent of the cosmic ray distribution is larger in the streaming case. Large heights reached by cosmic ray carrying gas in the non-streaming case can be attributed to the onset of star formation early on and the outflow is not sustainable in this case. On the other hand, the amount of cosmic ray energy carried away from the midplane is by comparison overwhelmingly large in the streaming case. Finally, the same trends are reflected in the distribution of the vertical component of the magnetic field (right column). As the simulations start from the initially purely toroidal fields, the presence of the vertical field component indicates that this field was generated due to non-negligible magnetic pressure, a Parker instability, and last but not least, due to the stellar feedback and cosmic ray streaming. Comparison of the non-streaming and streaming cases reveals that the latter case corresponds to much stronger vertical field component associated with a well developed outflow away from the disk.\\
\begin{figure*}
  \begin{center}
    \leavevmode
        \includegraphics[width=\textwidth]{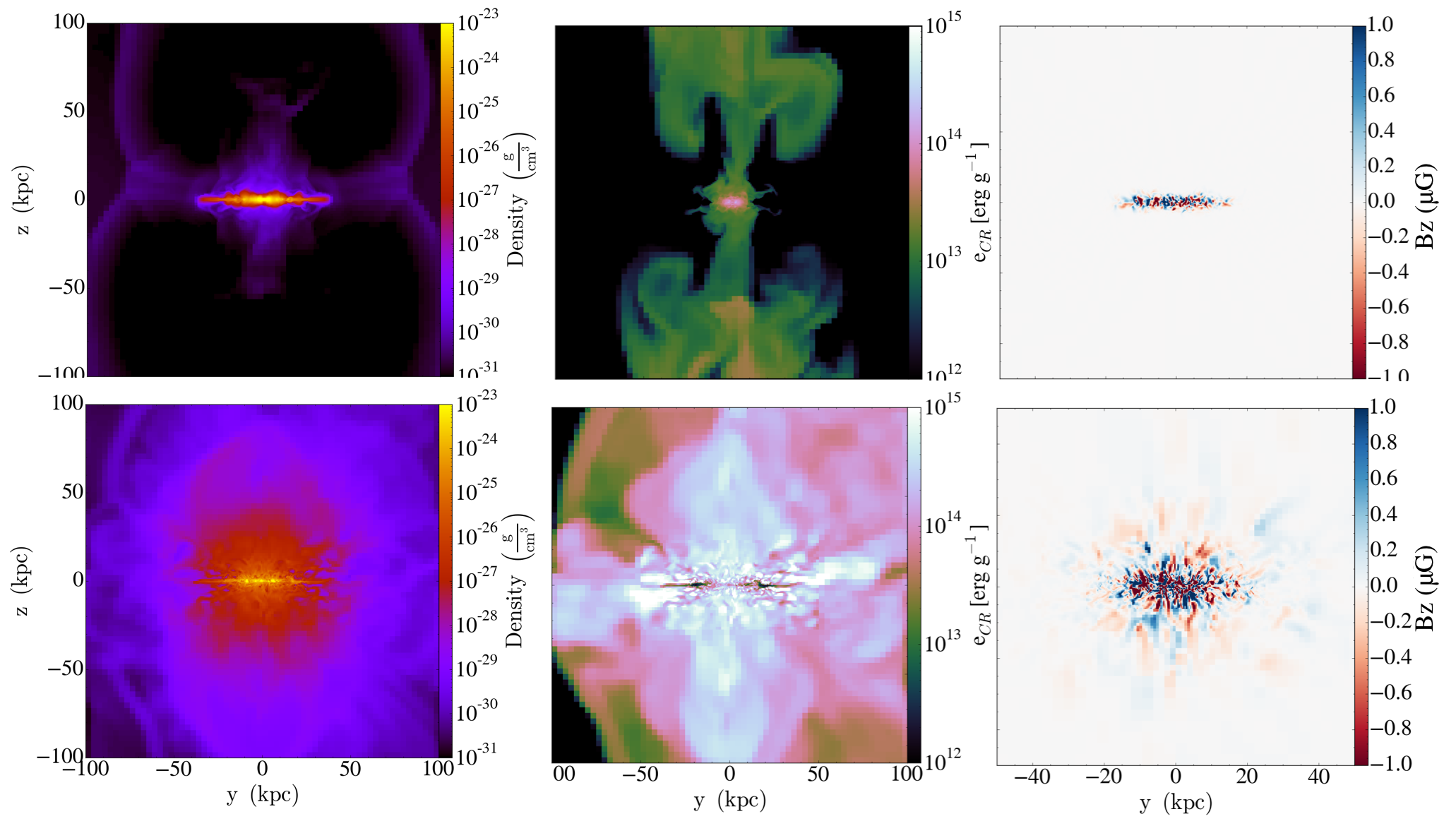}
       \caption[]{From left to right the panels show cross-sections through the distribution of gas density, specific energy density in CR, and the vertical component of the magnetic field. Top row corresponds to $f=0$ (i.e., no CR streaming) and bottom to $f=4$. All panels correspond to the SN feedback efficiency of 100$M_{\odot}$/SN, CR acceleration efficiency $f_{\rm cr}=0.1$, and no diffusion in either case. All snapshots were taken at 400 Myr since the beginning of the simulation. }
     \label{fig:dens}
  \end{center}
\end{figure*}
The impact of streaming on the wind launching discussed above is quantified in Figure 2. This figure shows the evolution of the mass outflow rate profiles for the cases presented in Figure 1. Panels show gas mass flux $\rho$u$_{z}$, where u$_{z}$ is the vertical component of the velocity field. The profiles were computed by weighting the mass flux by the gas cell mass in cylindrical bins along the $z$-direction. The weighting ensures that the results are not sensitive to the size of the volume within which the flux measurements are taken. The entire volume used in the computation of the profile was defined as a cylinder 50 kpc in radius and 100 kpc in height above the disk in the positive $z$-direction. The the cylinder base was located at the disk midplane. Such obtained profiles were then multiplied by 2 in order to account the mass flow in the top and bottom sides of the disk, and shown in Figure 2. The top panel corresponds to the non-streaming case and the bottom to the streaming case. In both panels the curves are plotted from 40 Myr to 450 Myr in increments of 10 Myr and are color-coded as indicated by the colorbar. The weak outflow seen in the non-streaming case (top panel) is not sustained over a long period of time. This initial outflow can be attributed to the transient star formation outburst in the beginning of the simulation. This has to be contrasted with the streaming case (bottom panel) where the outflow rate quickly increases and remains elevated over an extended period of time. These differences are not only evident in the sense of the magnitude of the mass flux, but are also reflected in the vertical extent of the region characterized by significant vertical mass flow. That is, toward the end of the simulation period shown in this figure, in the non-streaming case the extend of the region characterized by relatively large mass flux is confined to about a few kpc away from the disk, i.e., the disk essentially gently puffs up. This situation differs markedly from that seen in the streaming case, where the significant outflow extends to very large distances from the midplane of the galaxy. In this case, the gas is continuously accelerated as it is moving away from the disk.\\
\begin{figure}
  \begin{center}
    \leavevmode
        \includegraphics[width=\linewidth]{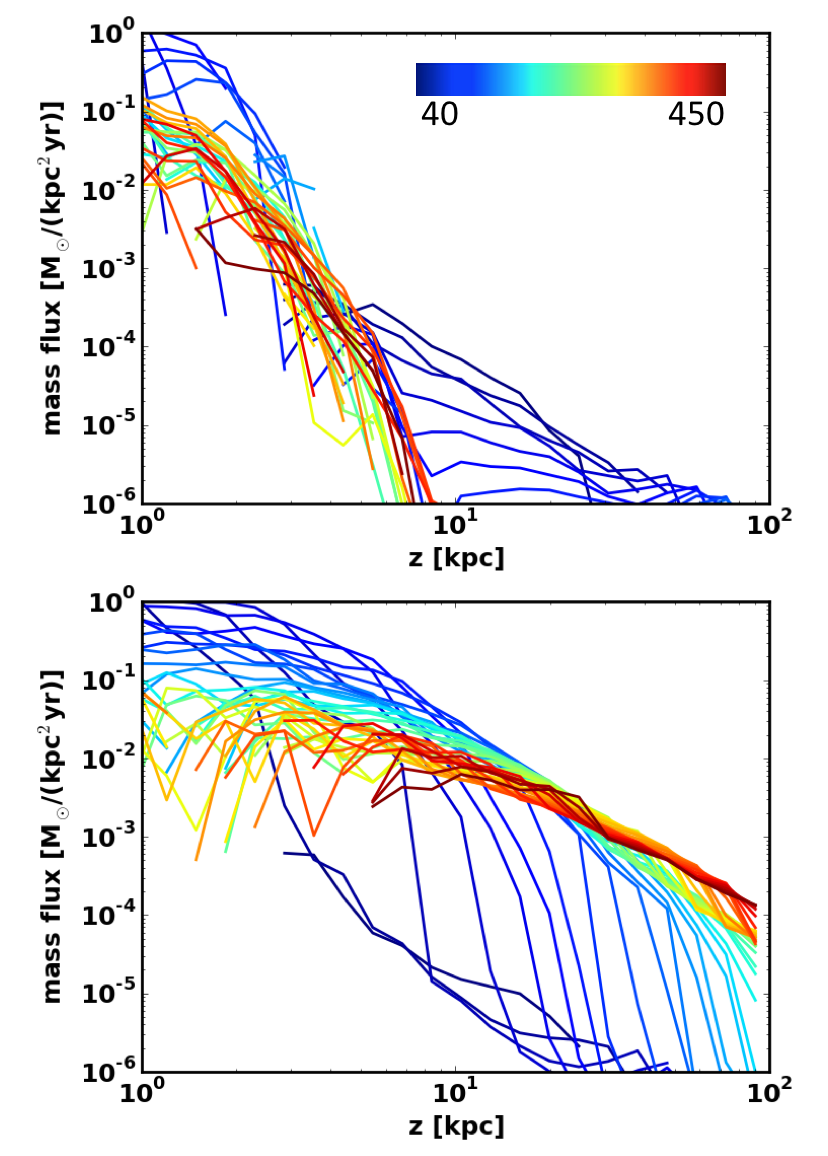}
       \caption[]{Evolution of the vertical distribution of the mass flux in the case without streaming ($f=0$; top panel) and including streaming ($f=4$; bottom panel). Both cases correspond to the SN feedback efficiency of 100$M_{\odot}$/SN and CR acceleration efficiency $f_{\rm cr}=0.1$. In each panel curves are plotted from 40 Myr to 450 Myr in increments of 10 Myr.}
     \label{fig:dens}
  \end{center}
\end{figure}
We now turn our attention to the reasons behind the differences between the simulations with and without streaming. Figure 3 presents cross-sections through the distribution of the vertical component of the magnetic field in the streaming case. The cross-section plane is perpendicular to the $x$ axis and positioned at $x=0$. Left panel shows the distribution close to the galactic mid-plane. Right panel presents a zoom-in on a part of the disk. The snapshot is taken at 400 Myr since the beginning of the simulation. Superimposed on these maps are the projected magnetic field vectors. The length of the vectors is proportional to the strength of the magnetic field. This figure demonstrates that while the magnetic field is highly tangled it also contains a partially ordered component in the form of ``chimneys'' extending from the disk midplane to larger distances below and above the disk. A comparison between this figure and the right column of Figure 1 reveals that such chimney structures are not seen in the non-streaming case. That is, while purely thermal driven chimneys are theoretically possible, they are not formed in our simulations. The formation of the chimneys is facilitated the uplift of the gas by SN and Parker instability aided by the streaming of CR along the magnetic fields. 
Using 2D MHD, and focusing on a small patch of a galactic disk, \citet{wang2010} demonstrated that anisotropic CR diffusion accelerates the development of Parker instability. They showed that for stronger diffusion, i.e., for weaker CR coupling to the gas, the magnetic field extends to larger heights above the disk. Our results are broadly consistent with their findings.
These structures allow the CR to stream out from denser regions of the disk and interact with more tenuous gas located just above the disk midplane. Cosmic ray pressure gradient acting on the gas with less inertia is more effective in accelerating it away from the disk midplane of the galaxy, thus launching the wind. The success or failure to launch the wind may depend on the detailed conditions close to the disk plane. \\
\begin{figure*}
  \begin{center}
    \leavevmode
        \includegraphics[width=\linewidth]{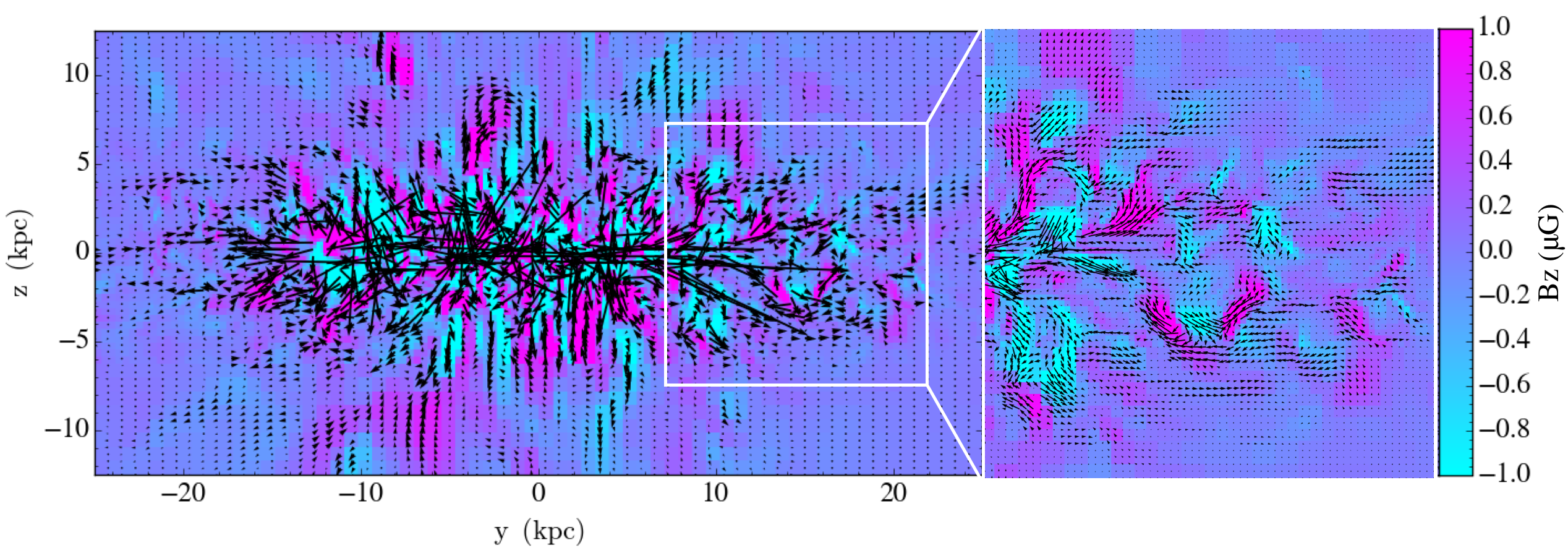}
       \caption[]{Cross-section through the distribution of the vertical component of the magnetic field (for the case corresponding to SN feedback efficiency of 100$M_{\odot}$/SN, CR acceleration efficiency $f_{\rm cr}=0.1$, and $f=4$.). Left panel shows the distribution close to the galactic mid-plane. Right panel presents a zoom-in on a part of the disk. The snapshot is taken at 400 Myr since the beginning of the simulation.}
     \label{fig:dens}
  \end{center}
\end{figure*}
As a next step we now explore in Figure 4 the impact of the various model parameters on the efficiency of wind driving and star formation in the disk. Galactic wind mass loading factors are shown in the top row and star formation rates are displayed in the bottom row. Mass loading factor is defined as the ratio of the baryon mass outside the cylindrical region 50 kpc in radius and 10 kpc in height centered on the midplane to the time-integrated star formation rate (which is the same as that used in \citet{salem2014} Bryan private comm.). \\
\indent
In the left column all cases correspond to $f=4$ and SN feedback efficiency of one SN per 100$M_{\odot}$ in newly formed stars. Other parameters are: $f_{\rm cr}=0.1$ (black); $f_{\rm cr}=0.15$ (red); $f_{\rm cr}=0.3$ (green); $f_{\rm cr}=0.15$, $B_{\rm o}=3\mu$G (blue).
The first three of these cases isolate the impact of the cosmic ray acceleration efficiency on the results. As expected, systematic increase in the acceleration efficiency leads to progressively larger mass loading factors in the saturated state, i.e., at late times. The mass loading factors range from $\sim 0.25$ to $\sim 0.6$. These values are broadly consistent with those obtained by \citet{booth2013} and \citet{salem2014} for Milky Way-like galaxy models (though the physics of transport processes is different in our case). The remaining case corresponds to increased initial magnetic field strength. In this case the effective streaming speed is larger (i.e., more CR can leak out of denser portions of the disk) which helps to drive the wind. However, the larger transport speed also results in shorter time during which CR energy density can build up in the disk as a result of SN feedback, which should reduce the force that CR exert on the gas to drive the wind. This case is further complicated by the fact that magnetic forces are stronger. The net outcome in this case is that the mass loading does increase early on but its final saturated value is only slightly larger than that corresponding to the lower magnetic field case. In terms of the dependence of the star formation rate on the cosmic ray acceleration efficiency, we find that star formation rates decrease as the efficiency increases. The suppression of star formation is quite significant, demonstrating that CR play a very important role in the stellar feedback process. \\
\indent
The middle column shows the dependence of the results on the effective streaming speed. In all cases $f_{\rm cr}=0.1$ and SN feedback efficiency of 100$M_{\odot}$ in newly formed stars per SN are chosen. Other parameters are $f=8$ (black), $f=4$ (red), $f=1$ (green), $f=0$ (blue). This column reveals a trend for the the mass loading to increase with the effective streaming speed. Note that the mass loading curve in the non-streaming case ($f=0$ case) in the top middle column is not shown due to the absence of the wind. The star formation 
also increases with increasing effective streaming speed. This trend can be understood by considering an extreme case of vanishing streaming speed. In such a case all CR that are produced remain in the disk and limit star formation but evidently do not increase the pressure in the disk enough to drive a wind. As CR leak out at progressively larger rates, the star formation is systematically less inhibited.\\
\indent
In the right column we present results for lower SN efficiency (185$M_{\odot}$/SN) and include two runs with anisotropic cosmic ray diffusion. We considered the following cases: $f=3$ (red), $f=1$ (green), $f=0$ (blue), $\kappa_{||}=10^{28}$cm$^{2}$s$^{-1}$ (no streaming; dashed), $\kappa_{||}=3\times 10^{27}$cm$^{2}$s$^{-1}$ (no streaming; dotted). All cases correspond to $f_{\rm cr}=0.1$. As in the cases shown in the middle column, here too we observe an increase in the mass loading factor with the effective streaming speed, and no wind in the non-streaming case. The mass loading factors for $f=1$ and two different SN efficiencies do not significantly differ at 500 Myr. However, the the differences are clear for larger streaming speeds (for $f=3$ and $f=4$), with larger SN efficiencies corresponding to larger mass loading factors (we attribute this difference to nearly twice the efficiency rather than a 25\% difference in $f$ between these two cases). This is consistent with the expectation that for vanishing (or suppressed) effective streaming speed, the wind (if any) should be relatively weak irrespectively of the amount of energy injection by SN. However, once CR are allowed to escape from the midplane, their ability to accelerate more tenuous gas is expected to be more pronounced the more CR are available, i.e., higher the SF feedback efficiency. This is further corroborated by other shorter test runs for larger $f$ (not shown), where we compared mass loading factors for two different SN efficiencies but identical $f$. As in the cases shown in the middle column, star formation rate in the right panel shows a monotonically increasing trend with the effective speed of cosmic ray streaming. Note that the non-streaming case is not shown in the upper panel of the rightmost column. This is simply because the wind does not develop in this case. \\
\indent
The set of curves shown in the right column also includes results corresponding to simulations where streaming is replaced by anisotropic cosmic ray diffusion. We considered two values of magnetic field aligned diffusion coefficient, $\kappa_{||}=10^{28}$cm$^{2}$s$^{-1}$ (dashed line) and $\kappa_{||}=3\times 10^{27}$cm$^{2}$s$^{-1}$ (dotted line), and observed that the mass loading increased with the level of diffusion. This trend may superficially appear to be in conflict with the results of \citet{salem2014}, who considered a range of values of isotropic diffusion coefficient, and found that the mass loading increases with decreasing diffusion. However, they also demonstrated that wind did not launch for vanishing diffusion. This argument shows that there is an optimal value of the effective diffusion where the effects of CR on wind driving are maximized. Below that critical effective isotropic diffusion $\kappa_{\rm crit}$ CR are efficiently trapped in the disk and are unable to overcome inertia of the dense gas near the midplane, and for diffusion significantly above $\kappa_{\rm crit}$ CR escape from the galaxy so fast that they do not have enough time to transfer enough momentum to the tenuous gas just above the midplane and to accelerate this gas. We suggest that \citet{salem2014} results correspond to the regime above $\kappa_{\rm crit}$, whereas our diffusion cases correspond to the opposite regime. This is consistent with the fact that the magnetic field near the midplane is highly tangled and so our anisotropic diffusivities essentially correspond to isotropic diffusivities $\kappa_{\rm iso}\sim \kappa_{||}/3$ that are comparable to or smaller than the smallest value $\kappa_{\rm iso}=3\times 10^{27}$cm$^{2}$s$^{-1}$ considered by \citet{salem2014}. In this scenario, our streaming cases correspond to the effective transport speed smaller than a hypothetical threshold beyond which streaming would be so fast as to render the wind driving inefficient. This is consistent with the fact that the separation between the asymptotic values of the mass loading decreases as the effective streaming speed increases.
However, the most efficient streaming case considered here falls short of that corresponding to the critical threshold and we do not study such extreme streaming cases as they are less physical. \\
\indent
Note also that the fiducial case considered by \citet{salem2014} can be reconciled with our results. Due to the significant tangling of the magnetic fields close to the disk, our run for $\kappa_{||}=10^{28}$cm$^{2}$s$^{-1}$ corresponds to the effective isotropic diffusivitiy $\kappa_{\rm iso} = 3\times 10^{27}$cm$^{2}$s$^{-1}$. In this case, the asymptotic value of the mass loading is $\sim 0.3$. This is the same level of mass loading as that obtained by \citet{salem2014}. Given that their fiducial case corresponds to $f_{\rm cr}=0.3$, while ours corresponds to $f_{\rm cr}=0.1$, and that their $\kappa_{\rm iso}=10^{28}$cm$^{2}$s$^{-1}$, while our $\kappa_{\rm iso}$ is 3 times lower, our results are not inconsistent if $\kappa_{\rm crit}\sim 3\times 10^{27}$cm$^{2}$s$^{-1}$.\\
\begin{figure*}
  \begin{center}
    \leavevmode
        \includegraphics[width=\linewidth]{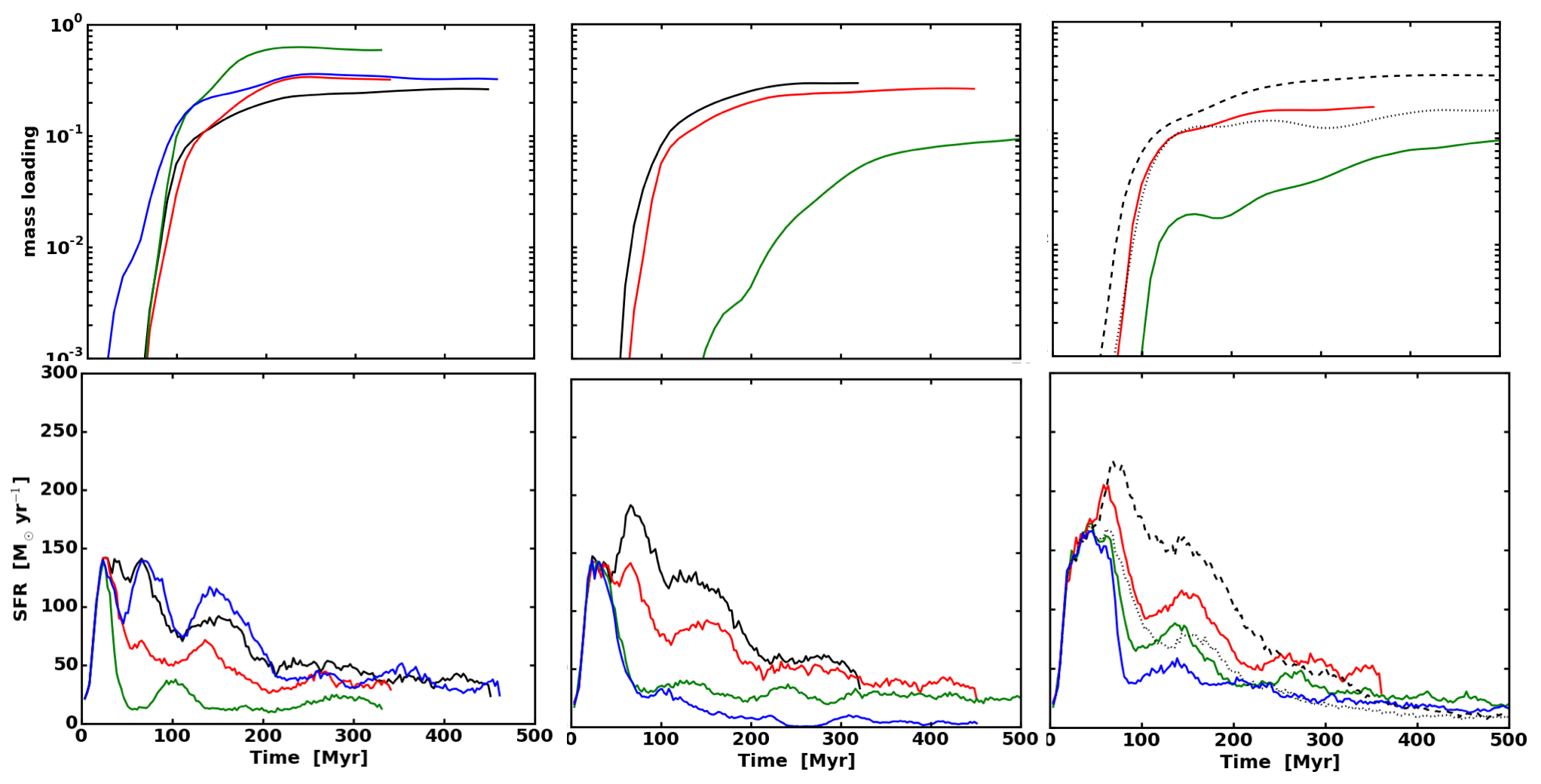}
       \caption[]{Galactic wind mass loading (top row) and star formation (bottom row). {\it Left} column (all cases for $f=4$, SN feedback efficiency of 100$M_{\odot}$/SN): $f_{\rm cr}=0.1$ (black); $f_{\rm cr}=0.15$ (red); $f_{\rm cr}=0.3$ (green); $f_{\rm cr}=0.15$, $B_{\rm o}=3\mu$G (blue); {\it Middle} column (all cases for $f_{\rm cr}=0.1$, SN feedback efficiency of 100$M_{\odot}$/SN): $f=8$ (black), $f=4$ (red), $f=1$ (green), $f=0$ (blue); {\it Right} column (all cases for $f_{\rm cr}=0.1$, SN feedback efficiency of 185$M_{\odot}$/SN): $f=3$ (red), $f=1$ (green), $f=0$ (blue), $\kappa_{||}=10^{28}$cm$^{2}$s$^{-1}$ (no streaming; dashed), $\kappa_{||}=3\times 10^{27}$cm$^{2}$s$^{-1}$ (no streaming; dotted). Note that the mass loading curves in the no-streaming cases ($f=0$ cases) in the middle and right columns are not shown due to the absence of the wind.}
     \label{fig:dens}
  \end{center}
\end{figure*}
\indent
We point out that the above values of the mass loading factors were obtained for specific parameters of the galaxy and, in particular, for one specific dark matter halo mass. Since this mass determines the depth of the gravitational potential well far from the disk, we expect that in smaller dark matter halos mass loading factors will be larger. This is consistent with  the findings of \citet{booth2013}. Most of the trends and dependencies on model parameters are likely to carry over to smaller galaxies as well. We intend to investigate and quantify these dependencies in detail in our subsequent work. We also note that the evolution of the mass loading factor factor would be different in live dark matter halo during the process of galaxy formation. As winds are observed directly at $z\sim 3$ (e.g., \citet{pettini2001, geach2005, law2007}), they propagate through significantly smaller dark matter halos (e.g., \citet{munozcuartas2011}). Consequently, the actual asymptotic values of the mass loading factors at $z=0$ (as defined here in the time-integrated sense) will be larger than those inferred without accounting for the growth of the dark matter halo, and it is the asymptotic levels of the mass loading that matter from the point of view of the total baryon budget.\\
\indent
While we intended to keep close to the model of \citet{salem2014}, we note that including an additional static potential, due to a preexisting stellar potential representing a part of the baryonic component, may facilitate easier wind launching. In this approach the total baryonic mass would be the same, but the amount of self-gravitating gas capable of forming new stars would be reduced. In this case, the gravitational field near the star formation sites is expected to be weaker and smoother, and thus CR do not need to accelerate gas from deeper local potential minima where stars form. We also note that the absolute level of mass loading may depend not only on including CR but also on the details of the modeling of the stellar feedback processes. These are important issues that deserve further investigation and we intend to address them, and the dependences of the results on halo/disk size, in a followup paper. \\

\begin{figure}
  \begin{center}
    \leavevmode
        \includegraphics[width=\linewidth]{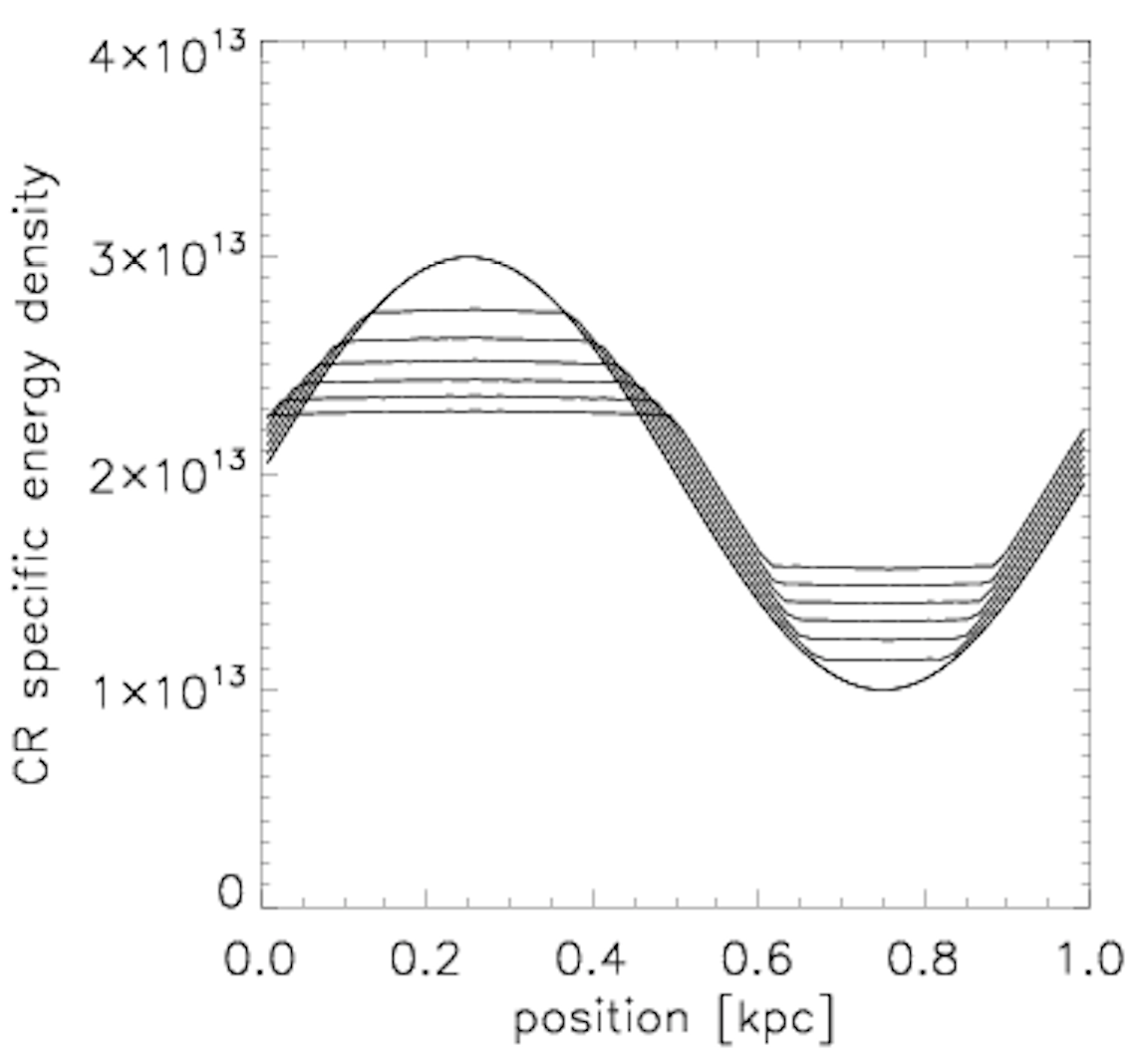}     
       \caption[]{The evolution of the specific energy density of CR due to cosmic ray streaming along magnetic fields. The initial conditions are described by a sine wave as a function of the $x$ coordinate. Streaming flattens the crest of the wave and fills the troughs. See Appendix for details.}
     \label{fig:ecr_mass}
  \end{center}
\end{figure}

\section{Summary and conclusions}
We have presented results from global three-dimensional simulations of galactic winds focusing on the dynamical role of CR and cosmic ray transport processes -- streaming and anisotropic diffusion. The main conclusions presented in this paper can be summarized as follows.

\begin{itemize}

\item We showed that including CR in the simulations significantly suppresses star formation rates. When no cosmic ray transport processes are included, the galactic disk puffs up but does not develop a sustained large-scale wind.

\item We found that efficient cosmic ray streaming and anisotropic diffusion can have a significant effect on the wind launching and mass loading factors depending on the details of the plasma physics. In particular, when the wave growth due to the streaming instability is inhibited by some damping process, such as the turbulent damping, the cosmic ray coupling to the gas is weaker and their effective propagation speed faster than the Alfv{\'e}n speed. We demonstrated that the presence of moderately super-Alfv{\'e}nic cosmic ray streaming enhances the efficiency of galactic wind driving. 

\item We demonstrated that CR considerably reduce the galactic star formation rates and significantly aid in launching galactic winds for cosmic ray acceleration efficiencies broadly consistent with the observational constraints.

\item In the cases where efficient transport effects operate, CR stream away from denser regions near the galactic disk along partially ordered magnetic fields and, in the process, accelerate more tenuous gas away from the galaxy. 

\end{itemize}

\acknowledgments{M.R. thanks Yuan Li, John Hewitt, Joel Bregman, Ryan Farber, and Brian O'Shea for useful discussions. H.Y.K.Y. acknowledges support by NASA through Einstein Postdoctoral Fellowship grant number PF4-150129 awarded by the Chandra X-ray Center, which is operated by the Smithsonian Astrophysical Observatory for NASA. The software used in this work was in part developed by the DOE NNSA-ASC OASCR Flash Center at the University of Chicago. M.R. acknowledges the NSF grant NSF 1008454 and NASA ATP 12-ATP12-0017. Simulations were performed on the Pleiades machine at NASA Ames. Data analysis presented in this paper was performed with the publicly available $yt$ visualization software \citep{turk2011}. We are grateful to the $yt$ development team and the $yt$ community for their support.\\}

\appendix
\section{Cosmic ray streaming module tests}
In order to test the CR streaming module we performed tests of CR streaming starting with sinusoidal initial conditions for the CR specific energy density. The wavelength of the initial perturbation is $\lambda = 1$ kpc. In this test we use $h_{\rm c}=10\lambda$ and constant background density $\rho = 10^{-24}$g cm$^{-3}$ and uniform magnetic field along the $x$-direction $B_{x}=3.545\times 10^{-6}$G. We assume periodic boundary condition in the $x$-direction. The evolution of this quantity due to cosmic ray streaming along magnetic fields is shown in Figure 5. The curves are plotted every 0.5 Myr. As expected, CR streaming flattens the crest of the wave and fills the troughs. While the parameters used for this test are more relevant to the galaxy wind simulations presented in this paper, we also performed tests for some of the same parameter choices as in \citet{sharma09} and obtained identical results.\\

\section{Connection between streaming and heating}
The energy equation for the thermal gas, which is derived from the first law of thermodynamics combined with the continuity equation, can be written in the form
\begin{equation}\label{energyt}
\frac{1}{\gamma_g - 1}\frac{\partial p_g}{\partial t} = -\bnabla\cdot\frac{p_g\mbfu_g}{\gamma_g-1} - p_g\bnabla\cdot\mbfu_g +\rho\frac{dQ}{dt},
\end{equation}
where we assumed the specific internal energy $e_{\rm th}$ is related to $p_g$ and $\rho$ by $p_g = (\gamma_g-1)\rho e_{\rm th}$ and only quantities which have a cosmic ray counterpart are subscripted. Let us assume the energy equation for the CR can be written in the form
\begin{equation}\label{energyc}
\frac{1}{\gamma_c - 1}\frac{\partial p_c}{\partial t} = -\bnabla\cdot\frac{p_c\mbfu_c}{\gamma_c-1} - p_c\bnabla\cdot\mbfu_c + \bnabla\cdot \left(
\bkappa\cdot\bnabla\frac{p_c}{\gamma_c - 1}\right).
\end{equation}
Equation (\ref{energyc}) can be derived from Eq. (9) of \citet{skilling1975}, which is a transport equation for the isotropic part of the cosmic ray distribution function, if we ignore second order Fermi acceleration (which is order $v_A^2/c^2$) and the transport velocity   and diffusion tensor $\bkappa$
are integrated over cosmic ray momentum in an appropriate way. This energy equation also agrees with Eq. (A11) of \citet{guooh2008}. The fluid momentum equation is the usual:
\begin{equation}\label{momentum}
\rho\frac{\partial\mbfu_g}{\partial t} = -\rho\mbfu_g\cdot\bnabla\mbfu_g-\bnabla\left(p_g+p_c\right).
\end{equation}
Let us define the energy density $W$ by
\begin{equation}\label{W}
W\equiv\frac{1}{2}\rho u_g^2+\frac{p_g}{\gamma_g-1}+\frac{p_c}{\gamma_c-1}.
\end{equation}

Using equations (\ref{energyt}), (\ref{energyc}), (\ref{momentum}), and the continuity equation leads to the following equation for the time
evolution of $W$:
\begin{equation}\label{Wdot}
\frac{\partial W}{\partial t}=-\bnabla\cdot\left[\frac{1}{2}\rho u_g^2\mbfu_g+\frac{\gamma_g p_g\mbfu_g}{\gamma_g-1}-\frac{\bkappa\cdot\bnabla p_c}{\gamma_c-1}\right]-
\frac{\gamma_c p_c\bnabla\cdot\mbfu_c}{\gamma_c-1}-\left(\mbfu_g+\frac{\mbfu_c}{\gamma_c-1}\right)\cdot\bnabla p_c+\rho\frac{dQ}{dt}.
\end{equation}
If we make the substitution
\begin{equation}\label{trick}
\mbfu_g+\frac{\mbfu_c}{\gamma_c-1} = \mbfu_c + (\mbfu_g-\mbfu_c) + \frac{\mbfu_c}{\gamma_c-1}  = (\mbfu_g-\mbfu_c) +
\frac{\gamma_c\mbfu_c}{\gamma_c-1}
\end{equation}
then Eq. (\ref{Wdot}) can be written as
\begin{equation}\label{fluxform}
\frac{\partial W}{\partial t}=-\bnabla\cdot\left[\frac{1}{2}\rho u_g^2\mbfu_g+\frac{\gamma_g p_g\mbfu_g}{\gamma_g-1}+\frac{\gamma_c p_c\mbfu_c}{\gamma_c-1}-
\frac{\bkappa\cdot\bnabla p_c}{\gamma_c-1}\right] +\rho\frac{dQ}{dt}+(\mbfu_c-\mbfu_g)\cdot\bnabla p_c.
\end{equation}
If we identify the heating rate as
\begin{equation}\label{heating}
\rho\frac{dQ}{dt} = -(\mbfu_c-\mbfu_g)\cdot\bnabla p_c
\end{equation}
then $\partial W/\partial t$ can be written as the divergence of a flux, which agrees with \citet{breitschwerdt1991} (exclusive of radiative
losses). For scattering by Alfv{\'e}n waves,  $\mbfu_c-\mbfu_g = \mbfu_A$ if all the waves propagate in the same direction as the cosmic
rays, but a more general definition is $\mbfu_c-\mbfu_g = f \mbfu_A$, where $f < 1$ is defined by Eq. (\ref{fdef}).
Note that streaming at effectively super-Alfv{\'e}nic speeds, which we parameterized by $f > 1$, does not result in a boost of the cosmic ray heating beyond $-\mbfu_A\cdot\bnabla p_c$, i.e., the heating term remains unchanged and is equal to $-\mbfu_A\cdot\bnabla p_c$.\ That
is because $f > 1$ accounts for the diffusive part of the cosmic ray flux when the marginal stability condition for the waves (super-Alfv{\'e}nic
streaming balances damping) is satisfied.\\

\bibliography{crwinds}

\end{document}